\documentclass[prl,twocolumn,10pt,aps,longbibliography,nofootinbib,superscriptaddress]{revtex4-1}

 \usepackage[utf8]{inputenc}
 \usepackage{bm}
 \usepackage{verbatim}
 \usepackage{amsmath}
 \usepackage{amssymb}
 \usepackage{amsthm}
 \usepackage{latexsym}
 \usepackage{amsfonts}
 \usepackage{epstopdf}
 \usepackage{wasysym} 
 \usepackage{subfigure}
 \usepackage{color}
 \definecolor{darkblue}{rgb}{0,0,.5}
 \definecolor{BLUE}{rgb}{0,0,1}
 \definecolor{BLACK}{rgb}{0,0,0}
 \usepackage[linktocpage, colorlinks=true, linkcolor=darkblue, citecolor=darkblue]{hyperref}
 \usepackage[all]{hypcap}
 \usepackage[makeroom]{cancel}
 \usepackage{xfrac}
 \usepackage{lipsum} 

\newcommand{\C}[1]{{\cal{#1}}}
\newcommand{\bb}[1]{\textbf{#1}}
\newcommand{\bs}[1]{\boldsymbol{#1}}

\newcommand{\lr}[1]{{\langle {#1}\rangle}}

\newcommand{\rl}[0]{{\rangle\langle}}

\begin{document}

\title{Typical Positivity of Nonequilibrium Entropy Production for Pure States}

\author{Philipp Strasberg}
\affiliation{Instituto de F\'isica de Cantabria (IFCA), Universidad de Cantabria--CSIC, 39005 Santander, Spain}
\affiliation{F\'isica Te\`orica: Informaci\'o i Fen\`omens Qu\`antics, Departament de F\'isica, Universitat Aut\`onoma de Barcelona, 08193 Bellaterra (Barcelona), Spain}
\author{Joseph Schindler}
\affiliation{F\'isica Te\`orica: Informaci\'o i Fen\`omens Qu\`antics, Departament de F\'isica, Universitat Aut\`onoma de Barcelona, 08193 Bellaterra (Barcelona), Spain}

\date{\today}

\begin{abstract}
 We establish that the nonequilibrium dynamics of most pure states gives rise to the same entropy production as that of the corresponding ensemble, provided the effective dimension of the ensemble is large enough. This establishes the positivity of entropy production under a wide variety of nonequilibrium situations. Our results follow from dynamical typicality and suitable continuity properties alone, without relying on non-integrability, and they complement other recent efforts to establish ``pure state second laws''. An explicit comparison with the distinctively different two-point measurement scheme is also provided.
\end{abstract}

\maketitle

\newtheorem{mydef}{Definition}[section]
\newtheorem{lemma}{Lemma}
\newtheorem{conj}{Conjecture}
\newtheorem{thm}{Theorem}[section]


Past decades have seen much progress in rigorously proving the non-negativity of entropy production (EP) for any Hamiltonian dynamics, even under nonequilibrium drivings and boundary conditions, and with associated fluctuation theorems for stochastic EP~\cite{EspositoHarbolaMukamelRMP2009, JarzynskiAnnuRevCondMat2011, CampisiHaenggiTalknerRMP2011, SagawaBook2012, BinderEtAlBook2018, DeffnerCampbellBook2019, StrasbergWinterPRXQ2021, StrasbergBook2022, PottsArXiv2024}. Yet, this could only be established by assuming the isolated Hamiltonian system to be prepared in a specific ensemble (usually related to a Gibbs state), making them unsatisfactory foundationally and also practically given recent experimental progress in quantum nanotechnologies. Below, we will call this the ``ensemble picture'' approach.

While numerical evidence suggests that the ensemble picture results also hold for pure states~\cite{SantosPolkovnikovRigolPRL2011, JinEtAlPRE2016, AlbaCalabresePNAS2017, SafranekDeutschAguirrePRA2019a, SafranekDeutschAguirrePRA2019b, KanekoIyodaSagawaPRE2019, FaiezEtAlPRA2020, HevelingEtAlPRE2022, PtaszynskiEspositoPRE2023, BabaYoshiokaSagawaArXiv2023, PandeyEtAlJSP2023, UsuiEtAlNJP2024, StrasbergSchindlerSP2024}, establishing this fact rigorously turned out to be arduous~\cite{VonNeumann1929, VonNeumannEPJH2010, VanKampenPhys1954, TasakiArXiv2000b, GoldsteinHaraTasakiArXiv2013, IyodaKanekoSagawaPRL2017, IyodaKanekoSagawaPRE2022, StrasbergEtAlPRA2023, NagasawaEtAlArXiv2024, HokkyoUedaArXiv2024}. Previous attempts relied on non-integrability~\cite{VonNeumann1929, VonNeumannEPJH2010, VanKampenPhys1954, TasakiArXiv2000b, GoldsteinHaraTasakiArXiv2013, IyodaKanekoSagawaPRL2017, IyodaKanekoSagawaPRE2022, StrasbergEtAlPRA2023, NagasawaEtAlArXiv2024, HokkyoUedaArXiv2024}, sometimes focused on a restrictive formulation of the second law~\cite{TasakiArXiv2000b, GoldsteinHaraTasakiArXiv2013, HokkyoUedaArXiv2024, IyodaKanekoSagawaPRL2017, IyodaKanekoSagawaPRE2022} (e.g., Planck's formulation), often assumed either the initial or final state to look equilibrated from a macroscopic point of view~\cite{VonNeumann1929, VonNeumannEPJH2010, TasakiArXiv2000b, GoldsteinHaraTasakiArXiv2013, IyodaKanekoSagawaPRE2022, NagasawaEtAlArXiv2024}, or used physical assumptions that are hard to prove microscopically~\cite{VanKampenPhys1954, StrasbergEtAlPRA2023} or were claimed invalid~\cite{IyodaKanekoSagawaPRL2017, GemmerKnipschildSteinigewegArXiv2017}. Remarkably, \emph{none} of these assumptions is necessary in the ensemble picture.

Here, we clarify this dissonance by showing that ``most'' pure quantum states give rise to an EP ``very close'' to the ensemble picture (in a sense made precise below) without any of the assumptions in Refs.~\cite{VonNeumann1929, VonNeumannEPJH2010, VanKampenPhys1954, TasakiArXiv2000b, GoldsteinHaraTasakiArXiv2013, IyodaKanekoSagawaPRL2017, IyodaKanekoSagawaPRE2022, StrasbergEtAlPRA2023, NagasawaEtAlArXiv2024, HokkyoUedaArXiv2024}. Our results rely on dynamical typicality~\cite{BartschGemmerPRL2009, MuellerGrossEisertCMP2011, MonnaiYuasaEPL2014, ReimannNC2016, BalzReimannPRL2017, ReimannPRE2018, DabelowReimannPRL2020, ReimannGemmerPA2020, DabelowReimannJSM2021, XuGuoPolettiPRA2022, TeufelTumulkaVogelJSP2023, TeufelTumulkaVogelArXiv2023b, TeufelTumulkaVogelAHP2024, ZhangEtAlNC2024} and certain continuity properties mentioned below. In particular, we will use a measure concentration result proven by Teufel, Tumulka and Vogel (Corollary 1 in Ref.~\cite{TeufelTumulkaVogelAHP2024}), which can be seen as a generalization of Levy's lemma~\cite{MilmanSchechtmanBook2001, PopescuShortWinterNatPhys2006} to non-uniform distributions.

\textbf{Lemma.} \emph{For any density matrix $\rho$ let $\mu_\rho$ be the Scrooge measure~\cite{JozsaRobbWoottersPRA1994} and let $B$ be a bounded operator. Then, for any $\epsilon\ge0$}
\begin{equation}\label{eq lemma}
 \mu_\rho\left\{\big|\lr{\psi|B|\psi} - \mbox{\normalfont tr}\{B\rho\}\big|>\epsilon\right\}
 \le 12 \exp\left(-\frac{C\epsilon^2}{\|B\|^2\|\rho\|}\right)
\end{equation}
\emph{with $C = (2304\pi^2)^{-1}$ a constant and $\|\cdot\|$ the operator norm.}

The Scrooge measure $\mu_\rho$ can be generated by sampling pure states $|\psi\rangle\sim\sqrt{\rho}|\phi\rangle$ with $|\phi\rangle$ a Haar random state, yielding $\rho = \lr{|\psi\rl\psi|}_\text{Scrooge}$ (with $\lr{\dots}_\text{Scrooge}$ the average over $\mu_\rho$). Given $\rho$, it is the most spread out ensemble because every (complete orthogonal rank-1) projective measurement yields the same information~\cite{JozsaRobbWoottersPRA1994}. It further arises naturally for open systems when sampling pure system-bath states in an suitable unbiased way~\cite{GoldsteinEtAlJSP2006, GoldsteinEtAlCMP2016}. Thus, since the Scrooge ensemble does not contain any information beyond $\rho$, it is natural to use it in the following to sample pure states. The smallness of the bound~(\ref{eq lemma}) is controlled by the smallness of $\|\rho\|$, which equals the maximum eigenvalue $\lambda_\text{max}$ of $\rho$. Its inverse $d_\text{eff} \equiv 1/\|\rho\|$ can be seen as an effective dimension of the system, describing how much $\rho$ is spread out over the Hilbert space; e.g., for a maximally mixed (microcanonical) state over some (sub)space $\C H$ of dimension $D$ one finds $d_\text{eff} = D$. This picture is also supported by noticing that $\|\rho\|^2 \le P(\rho) \le \|\rho\|$ with $P(\rho) = \mbox{tr}\{\rho^2\}$ the purity of $\rho$. Thus, $\|\rho\|$ is small, or spread out over a large $d_\text{eff}$, if and only if $\rho$ is highly mixed.

We now continue by analytically showing and numerically verifying our idea for the important setup of two finite systems exchanging energy, which has direct relevance, e.g., for transport experiments with cold atoms~\cite{BrantutEtAlScience2012, BrantutEtAlScience2013, HaeuslerEtAlPRX2021}. An explicit numerical comparison with the prominent two-point measurement scheme (TPMS)~\cite{EspositoHarbolaMukamelRMP2009, CampisiHaenggiTalknerRMP2011, DeffnerCampbellBook2019, StrasbergBook2022, PottsArXiv2024} is also given. Afterwards, we discuss possible caveats and ways to generalize the argument to other setups and sampling schemes. Another setup is also explicitly treated in the supplemental material~\cite{SupplementEntropyProductionAndDynamicalTypicality}.

\emph{Heat exchange setup.---}We consider a Hamiltonian $H=H_A+H_B+V$ describing two systems $A$ and $B$ coupled via an interaction $V$. The joint state of $AB$ evolves unitarily as $\rho(t) = e^{-iHt}\rho(0)e^{iHt}$ ($\hbar\equiv1)$. We introduce the following quantities for $A$ (and the same for $B$): $\rho_A(t) = \mbox{tr}_B\{\rho(t)\}$, $\pi_A(t) = e^{-\beta_A(t)H_A}/Z_A(t)$, $E_A(t) = \mbox{tr}_A\{H_A\rho_A(t)\}$, and $\C S_A(t) = -\mbox{tr}_A\{\pi_A(t)\ln\pi_A(t)\}$ are the reduced state, the Gibbs state at inverse temperature $\beta_A(t) = 1/T_A(t)$ ($k_B\equiv1$), the internal energy, and the thermodynamic entropy of $A$, respectively. More precisely, the thermodynamic (or canonical~\cite{SwendsenPRE2015, MattyEtAlPA2017, SeifertPA2020}) entropy is defined by choosing $\pi_A(t)$ to have the same energy as the exact time evolved state $\rho_A(t)$, i.e., $\mbox{tr}_A\{H_A\rho_A(t)\} = \mbox{tr}_A\{H_A\pi_A(t)\}$, which uniquely fixes $T_A(t)$. The particle number of system $A$ is denoted $N_A$.

Within the ensemble picture one assumes that the initial state $\rho(0) = \pi_A(0)\otimes \pi_B(0)$ is a product of two Gibbs states at (in general) different temperatures $T_A(0)\neq T_B(0)$. Note that this choice does \emph{not} imply $\rho_X(t)=\pi_X(t)$ (with $X\in\{A,B\}$) at later times $t>0$ owing to the interaction $V$. Under these conditions the following inequality was proven~\cite{StrasbergWinterPRXQ2021, StrasbergDiazRieraCampenyPRE2021}:
\begin{equation}\label{eq Clausius}
 \Sigma \equiv \sum_X\Delta\C S_X = \sum_X\int_0^\tau dt \frac{1}{T_X(t)}\frac{dE_X(t)}{dt} \ge 0.
\end{equation}
This is the expected second law from textbook thermodynamics under the assumption that only the energy (or temperature) of $A$ and $B$ is accessible. If the temperature $T_X(t)$ does not change appreciably during the interval $[0,\tau]$, Eq.~(\ref{eq Clausius}) transforms smoothly into the inequality $\sum_X\Delta E_X/T_X(0) \ge 0$, which has been derived independently~\cite{BassettPRA1978, JarzynskiWojcikPRL2004, EspositoLindenbergVandenBroeckNJP2010}.

For the pure state version of this second law we sample $|\psi(0)\rangle$ from the Scrooge measure $\mu_{\rho(0)}$ with $\rho(0) = \pi_A(0)\otimes\pi_B(0)$. Computing $d_\text{eff}^X = 1/\|\pi_X(0)\|$ is not possible without knowing $H$, but estimates are readily obtained. For instance, equivalence of ensembles implies that $\pi_X(0)$ is similar to a microcanonical state, whose $d_\text{eff}^X = \C O(10^N)$ scales exponentially with the particle number $N$. Likewise, if $X$ is composed of $N$ non-interacting (fermionic, bosonic, spin, etc.) modes $\mu$ each with purity $P(\pi_\mu)$, we find for the total purity $P[\pi_X(0)] = \prod_{\mu=1}^N P(\pi_\mu)$, which also has an exponential scaling with $N$. Thus, for instance for the cold atoms experiments of Refs.~\cite{BrantutEtAlScience2012, BrantutEtAlScience2013, HaeuslerEtAlPRX2021} one can reasonably assume $d_\text{eff}^X = \C O(10^{10000})$. Finally, quantities computed from $|\psi(t)\rangle = e^{-iHt}|\psi(0)\rangle$ get a superscript $\psi$ such as $E_X^\psi(t)$, $\C S_X^\psi(t)$ and $\beta_X^\psi(t)$ to distinguish them from the respective quantities appearing in the ensemble picture. The stochastic EP is then defined as
\begin{eqnarray}\label{eq stochastic EP Scrooge}
 \sigma_\text{can}^\psi \equiv \sum_X\Delta\C S_X^\psi,
\end{eqnarray}
where the subscript ``can'' refers to the direct use of the canonical entropy in computing Eq.~(\ref{eq stochastic EP Scrooge}). Note that the nonlinear dependence of $\C S_X^\psi(t)$ on $\psi$ implies $\Sigma\neq\lr{\sigma_\text{can}^\psi}_\text{Scrooge}$ in general. Nevertheless, $\sigma_\text{can}^\psi$ quantifies the EP in a single run of the experiment. In fact, time-of-flight measurements in cold atoms experiments~\cite{BlochDalibardZwergerRMP2008} reveal the temperature $T_X^\psi(t)$ from which $\C S_X^\psi(t)$ can be computed, whereas knowing the ensemble EP $\Sigma$ requires an unfeasible amount of experimental repetitions. Luckily, as we show now in three steps, the discrepancy between $\Sigma$ and $\sigma_\text{can}^\psi$ is negligible. This is the reason why thermodynamic experiments are reproducible~\cite{ReimannGemmerPA2020}, why the use of ensembles is justified in theoretical calculations~\cite{EspositoHarbolaMukamelRMP2009, JarzynskiAnnuRevCondMat2011, CampisiHaenggiTalknerRMP2011, SagawaBook2012, BinderEtAlBook2018, DeffnerCampbellBook2019, StrasbergWinterPRXQ2021, StrasbergBook2022, PottsArXiv2024}, and why one can use pure states to efficiently simulate ensemble dynamics~\cite{HeitmanEtAlZFN2020}.

First, we establish that the pure state energy $E_X^\psi(t)$ is close to the ensemble picture result for the overwhelming majority of states $|\psi(0)\rangle$ sampled from $\mu_{\rho(0)}$. This follows directly by choosing $B = U^\dagger(t) H_X U(t)$ in Eq.~(\ref{eq lemma}):
\begin{equation}\label{eq result 1}
 \begin{split}
  \mu_{\rho(0)}&\left\{\big|E_X^\psi(t) - E_X(t)\big|>N_X\epsilon\right\} \\
  & \le 12 \exp\left(-\frac{C\epsilon^2 N_X^2}{\|H_X\|^2} d_\text{eff}^A d_\text{eff}^B\right).
 \end{split}
\end{equation}
Here, we introduced the particle number $N_X$ to get the right scaling: since internal energy scales extensively, it matters that we can choose an $\epsilon\ll1$ such that the energy \emph{densities} are close with high probability. This is now guaranteed: since $\|H_X\|\sim N_X$, the right hand side scales as $e^{-d_\text{eff}^Ad_\text{eff}^B} \sim e^{-\exp(N_A+N_B)}$ with respect to $N_X$, thus strongly suppressing fluctuations away from the ensemble averaged density $E_X(t)/N_X$ even for moderate values of $N_X$. Importantly, Eq.~(\ref{eq result 1}) holds for all times $t$ (this is dynamical typicality).

Second, we establish that closeness in energies implies closeness in temperature. Using the heat capacity $C_X = dE_X/dT_X$ in the canonical ensemble, we write
\begin{equation}
 |E_X^\psi(t) - E_X(t)| = \left|\int_{T_X(t)}^{T_X^\psi(t)} C_X dT_X\right|.
\end{equation}
This integral is along a curve in temperature space and by the mean value theorem there exists a $\xi$ between $T_X^\psi(t)$ and $T_X(t)$ such that $|E_X^\psi(t) - E_X(t)| = C_X(\xi)|T_X^\psi(t) - T_X(t)|$. Assuming that the heat capacity is extensive, we set $c = C_X(\xi)/N_X$ and find
\begin{equation}\label{eq result 2}
 \begin{split}
 & \mu_{\rho(0)}\left\{\big|T_X^\psi(t) - T_X(t)\big|>\epsilon/c\right\} \\
 & = \mu_{\rho(0)}\left\{\big|E_X^\psi(t) - E_X(t)\big|>N_X\epsilon\right\},
\end{split}
\end{equation}
which establishes dynamical typicality of temperature. Again, this has been accomplished at the right scale: since temperature is intensive, no $N_X$-dependence appears in the first line. We remark that the bound becomes loose for $c\searrow0$ because the system then approaches zero temperature, which implies that the effective Hilbert space dimension $d_X$ is too small: too close to the ground state fluctuations dominate and typicality no longer applies.

Third and finally, we establish dynamical typicality of canonical entropy. To this end we note that 
\begin{equation}
 \begin{split}
  &|S_X^\psi(t) - S_X(t)| = \left|\int_{T_X(t)}^{T_X^\psi(t)} C_X \frac{dT_X}{T_X}\right| \\
  &= C_X(\xi')\left|\ln\frac{T_X^\psi(t)}{T_X(t)}\right| \le c'N_X \left|\frac{T_X^\psi(t)-T_X(t)}{T_X(t)}\right|,
 \end{split}
\end{equation}
where we again used the mean value theorem, set $c' = C_X(\xi')/N_X$, used $\ln(1+x)\le x$, and for simplicity considered the conventional case of positive temperatures $T_X>0$ (negative temperatures are better treated by using inverse temperatures, but the resulting expressions are more cumbersome and look less familiar). Combined with Eqs.~(\ref{eq result 1}) and~(\ref{eq result 2}) the above result implies
\begin{equation}\label{eq result 3}
\begin{split}
 \mu_{\rho(0)}&\left\{\big|S_X^\psi(t) - S_X(t)\big|>N_X\epsilon\frac{c'}{c}\frac{1}{T_X(t)}\right\} \\
& \le 12 \exp\left(-\frac{C\epsilon^2 N_X^2}{\|H_X\|^2} d_\text{eff}^A d_\text{eff}^B\right).
\end{split}
\end{equation}
We remark again that this establishes dynamical typicality of entropy at the right scale since entropy is extensive. Moreover, we see as before that typicality breaks down for $T_X(0)\searrow0$.

\begin{figure*}[t]
 \centering\includegraphics[width=0.99\textwidth,clip=true]{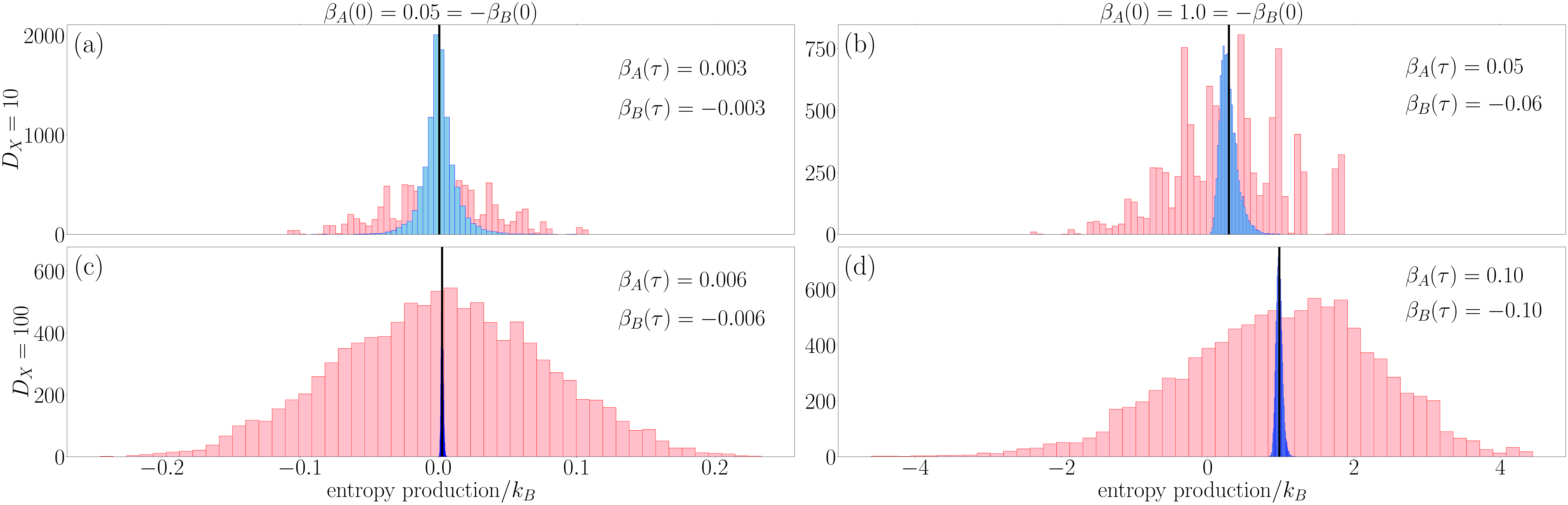}
 \label{fig numerics}
 \caption{Histograms of the stochastic EP for 10000 samples arranged into 50 bins for different initial conditions and system sizes (as explained in the main text). For a better direct comparison, the $x$ axis is shared in each column. }
\end{figure*}

To summarize, we find that fluctuations of $S_X^\psi(t)/N_X$ away from the ensemble picture result $S_X(t)/N_X$ are \emph{double} exponentially suppressed in $N_X$, which implies the same for the stochastic and ensemble picture EP $\sigma_\text{can}^\psi$ and $\Sigma$. This conclusion holds again with respect to the appropriate extensive scale $\sim N_X$. Possible caveats are discussed further below. 

\emph{Numerical demonstration.---}We check our results using a simple random matrix model. We use two identical Hamiltonians $H_A$ and $H_B$ with eigenenergies symmetrically drawn from a centered Gaussian density of states (which would arise, e.g., in spin systems) with unit variance. The subsystems are then weakly coupled with a banded random matrix $V$ with bandwidth $\delta_V=0.5$ and coupling strength $\lambda = (4D_A\delta_V)^{-1}$ in the notation of Ref.~\cite{StrasbergSchindlerSP2024} (the detailed construction is not necessary to understand the physical phenomenology; for similar models see also Refs.~\cite{BartschSteinigewegGemmerPRE2008, DasGhoshJSM2022, StrasbergReinhardSchindlerPRX2024}).

The blue histograms in Fig.~\ref{fig numerics} correspond to the stochastic EP $\sigma_\text{can}^\psi$. We consider a close-to-equilibrium condition with $\beta_A(0) = 0.05 = -\beta_B(0)$ and a far-from-equilibrium condition with $\beta_A(0) = 1.0 = -\beta_B(0)$ (left and right column of Fig.~\ref{fig numerics}, respectively). The local dimension of system $X$ is either $D_X = 10$ (top row) or $D_X = 100$ (bottom row), giving rise to a total dimension of $D_X^2=100$ or $D_X^2=10000$, respectively. Moreover, we let the system evolve for a nonequilibrium time scale $\tau=100\delta_V/\pi$~\cite{StrasbergSchindlerSP2024} after which global equilibrium is not yet reached, in particular for the far-from-equilibrium situation, as indicated by the final values of $\beta_X(\tau)$ in Fig.~\ref{fig numerics}. We observe that fluctuations around $\Sigma$ (black vertical lines) substantially decrease when increasing the Hilbert space dimension by a factor 100. Moreover, far from equilibrium no negative values for the EP are observed for all sampled states, indicating their extreme suppression. These conclusions confirm the theory above.

The pink histograms in Fig.~\ref{fig numerics} correspond to the distribution of stochastic EP within the TPMS. The TPMS samples energy eigenvalues $\bs\epsilon = (\epsilon_A^\tau,\epsilon_B^\tau,\epsilon_A^0,\epsilon_B^0)$ of $H_A$ and $H_B$ at the initial ($t=0$) and final ($t=\tau$) times, and the ``stochastic thermodynamics (st) definition'' of the fluctuating EP is 
\begin{eqnarray}\label{eq stochastic EP tpms}
 \sigma_\text{st}^{\bs\epsilon} \equiv \sum_X \left(-\ln\frac{e^{-\beta_X(\tau)\epsilon^X_\tau}}{Z_X(\tau)} + \ln\frac{e^{-\beta_X(0)\epsilon^X_0}}{Z_X(0)}\right).
\end{eqnarray}
Note that $\beta_X(t)$ and $Z_X(t)$ correspond to the Gibbs state $\pi_X(t)$ that matches the energy $E_X(t) = \mbox{tr}_X\{H_X\rho_X(t)\}$ of the \emph{ensemble picture} dynamics (it is \emph{not} the Gibbs state corresponding to the measured energy $\epsilon_X^t$). Therefore, the TPMS sampling and the definition of the stochastic EP are \emph{distinctively different} from the Scrooge sampling and definition~(\ref{eq stochastic EP Scrooge}). In particular,
\begin{enumerate}\setlength\itemsep{0em}
	\item $\Sigma = \lr{\sigma_\text{st}^{\bs\epsilon}}_\text{tpms}$ (with $\lr{\dots}_\text{tpms}$ the average over the TPMS sampling) whereas $\Sigma \neq \lr{\sigma_\text{can}^\psi}_\text{Scrooge}$,
	\item $\sigma_\text{st}^{\bs\epsilon}$ obeys an integral fluctuation theorem $\lr{\exp(-\sigma_\text{st}^{\bs\epsilon})}_\text{tpms} = 1$ (derived in~\cite{SupplementEntropyProductionAndDynamicalTypicality} by extending a trick of Tasaki~\cite{TasakiArXiv2000}) whereas $\sigma_\text{can}^\psi$ does not,
	\item $\sigma_\text{st}^{\bs\epsilon}$ can only be computed after knowing the ensemble averaged dynamics whereas $\sigma_\text{can}^\psi$ remains meaningful for a ``single shot'' experiment,
	\item $\sigma_\text{st}^{\bs\epsilon}$ requires to precisely measure microscopic energies whereas $\sigma_\text{can}^\psi$ only requires to measure the overall energy or temperature of $X$,
	\item it is unclear how to define $\sigma_\text{st}^{\bs\epsilon}$ if $\rho(0)$ has coherences (at least points 1.~and 2.~above can no longer both hold~\cite{PerarnauLlobetEtAlPRL2017}) whereas $\sigma_\text{can}^\psi$ remains well defined.
\end{enumerate}

In addition to these properties, Fig.~\ref{fig numerics} reveals that $\sigma_\text{st}^{\bs\epsilon}$ has much larger fluctuations than $\sigma_\text{can}^\psi$, which do not decrease with increasing Hilbert space dimension. Those are necessary to ensure the validity of the integral fluctuation theorem, which bounds the variance in EP only close to equilibrium~\cite{MerhavKafriJSM2010}. Even then fluctuations in $\sigma_\text{st}^{\bs\epsilon}$ exceed those in $\sigma_\text{can}^\psi$ by orders of magnitude [Fig.~\ref{fig numerics}(c)]. In fact, despite point 1.~above, we numerically observed that convergence to $\Sigma$ happens much faster for the Scrooge ensemble than for the TPMS for finite samples and large Hilbert spaces. Within the Scrooge ensemble a single trial thus typically encodes the full ensemble dynamics, in stark contrast to the conventional (quantum) stochastic thermodynamics approach.

Finally, one can mix the definitions above by combining the canonical entropy with the TPMS (resulting in giant fluctuations) or the stochastic thermodynamics entropy with the Scrooge ensemble (showing strong measure concentration). Explicit results are displayed in~\cite{SupplementEntropyProductionAndDynamicalTypicality}.

\emph{Generalizability.---}The above derivation is very general as it only relies on (i) dynamical typicality and (ii) certain continuity properties of the thermodynamic functions. 

The conditions for typicality are naturally satisfied in thermodyamics due to coarse-graining and large system sizes (with possible caveats very close to zero temperature).

Point (ii) is more subtle as entropy and temperature are complicated functions. For instance, above we assumed that the heat capacity densities $c$ and $c'$ are independent of $N_X$, but they can diverge at phase transitions (e.g., in the classical $2D$ Ising model~\cite{OnsagerPR1944, FerdinandFisherPR1969}), which \emph{could} jeopardize the argument. Thus, critical points require more care and non-analytical behaviour of the EP has been observed~\cite{ZhangBaratoJSM2016, BarbosaTomeJPA2019, VariziPRR2020, GoesFioreLandiPRR2020, MartynecKlappLoosNJP2020, CaballeroCatesPRL2020}. Moreover, quantities such as Shannon entropy are not Lipschitz continuous at the extreme points of the convex state space. Whether this causes troubles depends on the initial ensemble and is also a matter of scale. For instance, in the above heat exchange setup we could have included a small system $S$ in between $A$ and $B$ (a conventional scenario in quantum transport~\cite{NazarovBlanterBook2009, SchallerBook2014, SothmannSanchezJordanNanotechnology2015, BenentiEtAlPhysRep2017}). Now, suppose that $S$ has $d_S$ many levels. Then, its contribution to the EP is bounded by $\ln(d_S)$, whereas the heat exchanged between the reservoirs $A$ and $B$ grows indefinitely with time $t$ until global equilibrium is reached. Since usually $d_A,d_B\gg d_S$ the non-analyticity of Shannon (or von Neumann) entropy of the system $S$ might not matter for many relevant scenarios. In particular, it does \emph{not} matter for steady state transport.

Finally, whether most pure states have a \emph{positive} entropy production also depends on the scale of $\Sigma$. For instance, above $\Sigma$ would be very small if $A$ and $B$ were coupled only for a femtosecond such that they could barely exchange any energy quanta. Even if dynamical typicality would hold, positivity of EP could not be guaranteed for pure states with high probability. This is analogous to the equilibrium case with $\Sigma=0$: fluctuations with positive and negative EP must (by definition of equilibrium) be symmetrically distributed. Only far away from equilibrium (for large $\Sigma$) is it possible to establish that most pure states comply with the arrow of time.

Despite these subtleties, the above reasoning is remarkably robust and applies to a broad class of situations. Special care is only required for small effective Hilbert space dimension $d_X$ (e.g., at very low temperatures), when very fine-grained information is available, or at criticality. In particular, ``special care'' does not imply that the here presented argument is inapplicable (for instance, while the heat capacity density might diverge so does $d_X$ in the exponent too). However, these details together with the fact that there is (up to now) no ``master theorem'' for EP unifying all the frameworks of Refs.~\cite{EspositoHarbolaMukamelRMP2009, JarzynskiAnnuRevCondMat2011, CampisiHaenggiTalknerRMP2011, SagawaBook2012, BinderEtAlBook2018, DeffnerCampbellBook2019, StrasbergWinterPRXQ2021, StrasbergBook2022, PottsArXiv2024} forbids us to present our argument as a simple, universally applicable theorem. This might simply accurately reflect the complexity of statistical physics, but it is worth to search for such a master theorem, for instance, using a unified framework of thermodynamic entropies~\cite{SchindlerEtAlArXiv2024}. To illustrate some of the features above, we treat another example (a driven isolated system) in~\cite{SupplementEntropyProductionAndDynamicalTypicality} using observational entropy~\cite{StrasbergWinterPRXQ2021, SafranekEtFP2021} and its continuity properties~\cite{SchindlerWinterJMP2023}.

\emph{Other samplings.---}While it seems natural to use the Scrooge ensemble as the most unbiased ensemble given $\rho$, other samplings are conceivable too. Above, we already studied the TPMS, which did \emph{not} give rise to measure concentration for EP. However, many other reasonable ensembles would do. For instance, we could sample states $|\psi(0)\rangle$ with \emph{fixed} energy $E_X^\psi(0) = E_X(0)$ as considered by M\"uller, Gross and Eisert, who also prove a measure concentration result~\cite{MuellerGrossEisertCMP2011}. As Reimann has shown in a detailed calculation, the resulting implications for dynamical typicality are qualitatively the same~\cite{ReimannPRE2018}. Intuitively, this makes sense because most states $|\psi(0)\rangle$ in the Scrooge ensemble have an energy extremely close to the average [see Eq.~(\ref{eq result 1})] and, by continuity, one expects that the majority of states with exactly the same energy behaves similarly to the states in a small $\epsilon$-shell around it. Alternatively, we could have sampled initial states from microcanonical energy shells in $A$ and $B$. In this case we could apply Levy's lemma~\cite{MilmanSchechtmanBook2001, PopescuShortWinterNatPhys2006} to again establish dynamical typicality of the entropy production (or we could invoke equivalence of ensembles).

\emph{Conclusions.---}Using dynamical typicality, we established a robust and general strategy to derive pure state second laws that complements existing approaches~\cite{VonNeumann1929, VonNeumannEPJH2010, VanKampenPhys1954, TasakiArXiv2000b, GoldsteinHaraTasakiArXiv2013, IyodaKanekoSagawaPRL2017, IyodaKanekoSagawaPRE2022, StrasbergEtAlPRA2023, NagasawaEtAlArXiv2024, HokkyoUedaArXiv2024}: no assumption about non-integrability was needed, and both the initial and final states could be far from equilibrium (i.e., the present approach is applicable to transient dynamics and nonequilibrium steady states).

Our result closes a gap by curing deficiencies of both the ensemble picture and previous pure state second laws. In fact, while the ensemble picture proves $\Sigma\ge 0$, it does not prove that entropy gets maximized and it does not exclude arbitrary strong and frequent recurrences down to $\Sigma = 0$. Conversely, while von Neumann's $H$-theorem~\cite{VonNeumann1929, VonNeumannEPJH2010} proves the eventual maximization of entropy, it does not allow conclusions about transient dynamics. Our result implies that transient second-law-violating states are exceptionally rare (but in unison with time reversal symmetry not impossible), and it suggests that temporary violations of the second law, or recurrences of the EP, are unlikely unless the dynamics guides the state into atypical second-law-violating subspaces. This behaviour cannot be excluded in the ensemble picture because a repeated use of the Gibbs state assumption would violate unitarity of the dynamics, but---as we showed---it only matters that the pure state is sufficiently representative of a Gibbs state at later times, see also~\cite{VanKampenPhys1954, StrasbergEtAlPRA2023}.

Our approach also sheds light on contemporary debates in quantum thermodynamics. While the ensemble picture seemingly provides a consistent thermodynamic framework for any bath size, our results indicate that a good bath needs a minimum size $d_\text{eff}$: if the bath is too small, the thermodynamics sensitively depends on inaccessible information about the precise state of the bath. Somewhat echoing this contrast, a definition of EP has been suggested that diverges if the bath is in a pure state~\cite{LandiPaternostroRMP2021} or that a pure state bath has zero temperature~\cite{ElouardLombardLatunePRXQ2023}. This conflicts with our approach and also with other recent findings~\cite{ODonovanArXiv2024}.

Finally, the present approach is not restricted to weak coupling, can be applied to multiple conserved charges (including non-commuting ones~\cite{MajidyEtAlNRP2023}), might provide an experimentally more meaningful alternative to the TPMS with its unrealistic perfect energy resolution (for other alternatives see, e.g., Refs.~\cite{AllahverdyanNieuwenhuizenPRE2005, MillerAndersNJP2017, StrasbergPRE2019, HamedMohammadyRomitoQuantum2019, MicadeiLandiLutzPRL2020, KerremansSamuelssonPottsSPP2022}), and therefore provides a useful tool for many current research directions.

\emph{Acknowledgements.---}We gratefully acknowledge discussions with Roderich Tumulka and Andreas Winter. Finanical support by MICINN with funding from European Union NextGenerationEU (PRTR-C17.I1) and by the Generalitat de Catalunya (project 2017-SGR-1127) are acknowledged. PS is further supported by ``la Caixa'' Foundation (ID 100010434, fellowship code LCF/BQ/PR21/11840014), the Ram\'on y Cajal program RYC2022-035908-I, the European Commission QuantERA grant ExTRaQT (Spanish MICIN project PCI2022-132965), and the Spanish MINECO (project PID2019-107609GB-I00) with the support of FEDER funds.


\bibliography{/home/philipp/Documents/references/books.bib,/home/philipp/Documents/references/open_systems.bib,/home/philipp/Documents/references/thermo.bib,/home/philipp/Documents/references/info_thermo.bib,/home/philipp/Documents/references/general_QM.bib,/home/philipp/Documents/references/math_phys.bib,/home/philipp/Documents/references/equilibration.bib,/home/philipp/Documents/references/time.bib,/home/philipp/Documents/references/cosmology.bib,/home/philipp/Documents/references/general_refs.bib}

\begin{thebibliography}{96}
\bibitem [{Sup()}]{SupplementEntropyProductionAndDynamicalTypicality}%
  \BibitemOpen
  \href@noop {} {}\bibinfo {note} {The SM contains a derivation of an integral
  fluctuation theorem, it numerically compares further definitions and
  samplinges of the stochastic EP, and it derives dynamical typicality of EP
  for another example in detail.}\BibitemShut {Stop}%
\end{thebibliography}

\appendix
\section{Supplemental material}

\subsubsection{Integral fluctuation theorem for Clausius' inequality}

Let $H_X = \sum_\epsilon \epsilon^X\Pi(\epsilon^X)$ be the spectral decomposition of the Hamiltonian of $X$ and consider the probability to measure energies $\epsilon_0^X$ at time $t=0$ and $\epsilon_\tau^X$ at time $t=\tau$:
\begin{equation}
 p(\bs\epsilon) =
 \mbox{tr}\{\Pi(\bs\epsilon_\tau) e^{-iH\tau} \Pi(\bs\epsilon_0) \rho(0) \Pi(\bs\epsilon_0) e^{iH\tau}\}.
\end{equation}
Here, we used the concise notation $\bs\epsilon = (\epsilon_A^\tau,\epsilon_B^\tau,\epsilon_A^0,\epsilon_B^0)$ and $\Pi(\bs\epsilon_t) = \Pi(\epsilon_t^A)\Pi(\epsilon_t^B)$. Next, we define the stochastic entropy of system $X$ conditional on measuring $\epsilon^X_t$ ($t\in\{0,\tau\}$) as
\begin{equation}
 s_X^{\epsilon^X_\tau,\epsilon^X_0} = -\ln\frac{e^{-\beta_X(t)\epsilon_t^X}}{Z_X(t)},
\end{equation}
where $\beta_X(t)$ and $Z_X(t) = \mbox{tr}_X\{\pi_X(t)\}$ are the inverse temperature and the partition function of the Gibbs state $\pi_X(t)$ associated to the ensemble picture dynamics, defined via the relation $\mbox{tr}_X\{H_X\rho_X(t)\} = \mbox{tr}_X\{H_X\pi_X(t)\}$. Consequently, the stochastic EP is $\sigma_\text{st}^{\bs\epsilon} = \sum_X\Delta s_X$ [Eq.~(\ref{eq stochastic EP tpms}) in the main text]. The probability distribution $p(\sigma)$ becomes
\begin{equation}
 p(\sigma) =
 \sum_{\bs\epsilon} \delta\left(\sigma-\sigma_\text{st}^{\bs\epsilon}\right) p(\bs\epsilon),
\end{equation}
where $\delta(\cdot)$ denotes the delta function.

Now, choosing as the initial state $\rho(0) = \pi_A(0)\otimes\pi_B(0)$ as in the main text, we find
\begin{equation}
 \begin{split}
  p&(\bs\epsilon) = \\
  & \exp\left(-\sum_X s_X^{\epsilon^X_0}\right) \mbox{tr}\{\Pi(\bs\epsilon_\tau) e^{-iH\tau} \Pi(\bs\epsilon_0) e^{iH\tau}\}.
 \end{split}
\end{equation}
Using the property of the delta function, we find
\begin{align}
 \lr{e^{-\sigma}} &= \int d\sigma e^{-\sigma} p(\sigma) \\
 &= \sum_{\bs\epsilon} \exp\left(-\sum_X s_X^{\epsilon^X_\tau}\right) \mbox{tr}\{\Pi(\bs\epsilon_\tau) e^{-iH\tau} \Pi(\bs\epsilon_0) e^{iH\tau}\} \nonumber \\
 &= 1, \nonumber
\end{align}
where the last line follows from the completeness of the projectors $\Pi(\bs\epsilon_0)$, unitarity, and normalization.

We remark that the mathematical idenity $\lr{e^{-\sigma}} = 1$ holds for \emph{any} choice of $\beta_X(\tau)$. However, only if we choose $\beta_X(\tau)$ as the inverse temperature matching the final average energy (as in the main text), we find that $\Sigma = \lr{\sigma_\text{st}^{\bs\epsilon}}_\text{tpms}$. This has been noticed by Tasaki for a single isolated system~\cite{TasakiArXiv2000}. Choosing instead $\beta_X(\tau) = \beta_X(0)$ yields the conventional exchange fluctuation theorem (see, e.g., Refs.~\cite{JarzynskiWojcikPRL2004, EspositoHarbolaMukamelRMP2009, CampisiHaenggiTalknerRMP2011, StrasbergBook2022, PottsArXiv2024}).

\begin{figure*}[t]
	\centering\includegraphics[width=0.99\textwidth,clip=true]{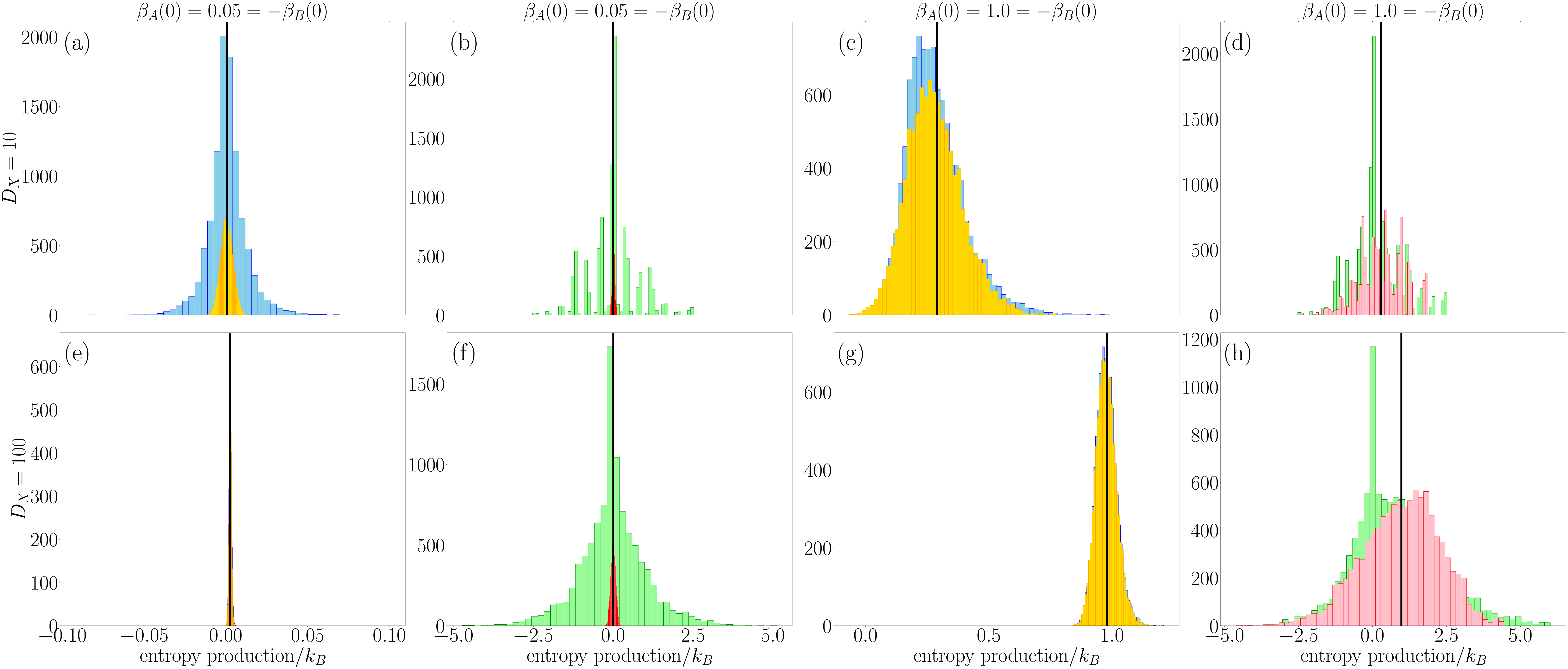}
	\label{fig numerics new}
	\caption{Histograms of the stochastic EP for 10000 samples arranged into 50 bins for different initial conditions and system sizes (as explained in the main text). For a better direct comparison, the $x$ axis is shared in each column. }
\end{figure*}

\subsubsection{Other samplings and definitions}

As remarked in the main text, there are more than two possibilities to define fluctuations in a stochastic EP. In particular, we studied two different sampling schemes (Scrooge ensemble versus TPMS) and two possible definitions of stochastic EP (direct evaluation of the change in canonical entropy versus the stochastic thermodynamics approach, which requires prior knowledge of the ensemble). We can summarize the resulting four definitions as:
\begin{align}
 \sigma_\text{can}^\psi &\equiv \sum_X \left\{\C S_X[E_X^\psi(\tau)] - \C S_X[E_X^\psi(0)]\right\}, \\
 \sigma_\text{st}^\psi &\equiv \sum_X \left(-\ln\frac{e^{-\beta_X(\tau)E_X^\psi(\tau)}}{Z_X(\tau)} + \ln\frac{e^{-\beta_X(0)E_X^\psi(0)}}{Z_X(0)}\right), \\
 \sigma_\text{can}^{\bs\epsilon} &\equiv \sum_X \left\{\C S_X[\epsilon_\tau^X] - \C S_X[\epsilon_0^X]\right\}, \\
 \sigma_\text{st}^{\bs\epsilon} &\equiv \sum_X \left(-\ln\frac{e^{-\beta_X(\tau)\epsilon^X_\tau}}{Z_X(\tau)} + \ln\frac{e^{-\beta_X(0)\epsilon^X_0}}{Z_X(0)}\right).
\end{align}
The resulting histograms are displayed in Fig.~\ref{fig numerics new} for the same system as studied in the main text. Again, the upper row refers to $D_X=10$ and the lower row to $D_X=100$. However, due to the many histograms and the vast differences in scale, we decided to plot $\sigma_\text{can}^\psi$ (blue bars as in the main text) and $\sigma^\psi_\text{st}$ (golden bars) together but separate from $\sigma_\text{st}^{\bs\epsilon}$ (pink bars as in the main text) and $\sigma_\text{can}^{\bs\epsilon}$ (green bars). Note the difference in scale of the $x$ axes between the first and second or third and fourth column. In particular, close to equilibrium [Fig.~\ref{fig numerics new}(b) and (f)] $\sigma_\text{can}^{\bs\epsilon}$ shows enormous fluctuations. In contrast, $\sigma_\text{can}^\psi$ and $\sigma^\psi_\text{st}$ behave quite similarly, which follows from the fact that both the energy and canonical entropy obey dynamical typicality as shown in the main text.

\subsubsection{Dynamical typicality of EP for driven systems}

We here establish dynamical typicality of EP for another setting and using a different entropy notion; namely, observational entropy instead of canonical entropy (however, it has been recently established that both can be treated within a unifying framework~\cite{SchindlerEtAlArXiv2024}).

We consider a driven isolated system with Hamiltonian $H(\lambda_t)$, where $\lambda_t$ denotes some externally prescribed driving protocol. We remark that the isolated system could be split into subsystems (e.g., a system and bath), but this is irrelevant for our discussion. In terms of its instantaneous energy eigenvectors and eigenvalues we write $H(\lambda_t) = \sum_{i=1}^D \epsilon_i(\lambda_t)|i(\lambda_t)\rl i(\lambda_t)|$. Now, we assume that we only know the energy up to some uncertainty (or measurement error) $\delta$. We introduce projectors
\begin{equation}
 \Pi_E = \sum_{i\in I_E} |i(\lambda_t)\rl i(\lambda_t)|
\end{equation}
associated to an energy interval $I_E = \{i|E\le\epsilon_i< E+\delta\}$ that add up to the identity, $\sum_{E=1}^M \Pi_E = I$ (we assume there are $M$ projectors), and are orthogonal $\Pi_E\Pi_{E'} = \delta_{E,E'}\Pi_E$. Note that all the previous quantities depend on $\lambda_t$, which we suppress in the notation for conciseness.

If $\rho(t)$ is the state of the system and $p_E(t) = \mbox{tr}\{\Pi_E\rho(t)\}$ the probability to find the system with (coarse) energy $E$, the associated observational entropy is~\cite{SafranekEtFP2021, StrasbergWinterPRXQ2021}
\begin{equation}
 S_E(t) = \sum_E p_E(t)[-\ln p_E(t) - \ln V_E],
\end{equation}
where $V_E = \mbox{tr}\{\Pi_E\}$ is the Boltzmann volume associated to the subspace with measurement outcome $E$. Within the ensemble picture it has been established that for any initial state of the form
\begin{equation}\label{eq initial state 2}
 \rho(0) = \sum_E p_E(0) \frac{\Pi_E}{V_E}
\end{equation}
with an arbitrary initial distribution $p_E(0)$ the following second law holds~\cite{StrasbergWinterPRXQ2021}
\begin{equation}\label{eq EP}
 S_E(\tau) - S_E(0) \ge 0,
\end{equation}
which describes the increase in entropy in the isolated system for any $H(\lambda_t)$.

To establish dynamical typicality of EP, we proceed as in the main text. First of all, a straightforward application of Eq.~(\ref{eq lemma}) gives
\begin{equation}\label{eq result 1 2}
 \mu_{\rho(0)}\left\{\big|p_E^\psi(t) - p_E(t)\big|>\epsilon\right\} \le 12 \exp\left(-\frac{C\epsilon^2}{\|\rho(0)\|}\right).
\end{equation}
Here, as in the main text, we denoted $p_E^\psi(t) = \lr{\psi(t)|\Pi_E|\psi(t)}$ with $|\psi(0)\rangle$ drawn from the Scrooge ensemble corresponding to Eq.~(\ref{eq initial state 2}). Note that the bound on the right hand side is independent of $E$ and $t$.

For a coarse measurement (sufficiently small $M$) and a sufficiently spread out ensemble $\rho(0)$, this implies that the distributions $\bb p(t)$ and $\bb p^\psi(t)$ must be close in trace norm
\begin{equation}
 \begin{split}
  \Delta &\equiv \frac{1}{2}\|\bb p(t) - \bb p^\psi(t)\|_1 = \frac{1}{2}\sum_E |p_E^\psi(t) - p_E(t)| \\
  &\le \frac{M}{2} \max_E\{|p_E^\psi(t) - p_E(t)|\}
 \end{split}
\end{equation}
because the last inequality implies
\begin{align}\label{eq result 2 2}
 \mu_{\rho(0)}\left\{\Delta>\epsilon\right\}
 &\le \mu_{\rho(0)}\left\{\max_E\big|p_E^\psi(t) - p_E(t)\big|>\frac{2\epsilon}{M}\right\} \nonumber \\
 &\le 12 \exp\left(-\frac{4C\epsilon^2}{M^2\|\rho(0)\|}\right).
\end{align}

Finally, we use the following continuity bound for observational entropy~\cite{SchindlerWinterJMP2023}: for a given trace norm $\Delta$
\begin{equation}
 |S_E^\psi(t)-S_E(t)| \le g(\Delta) + \Delta\ln D
\end{equation}
with $g(x) = -x\ln x+(1+x)\ln(1+x)$ and $D$ the Hilbert space dimension. It follows that
\begin{equation}
 \begin{split}
  & \mu_{\rho(0)}\left\{|S_E^\psi(t)-S_E(t)| > g(\epsilon) + \epsilon\ln D \right\} \\
  &\le \mu_{\rho(0)}\{\Delta > \epsilon\} \le 12 \exp\left(-\frac{4C\epsilon^2}{M^2\|\rho(0)\|}\right).
 \end{split}
\end{equation}

To demonstrate that this bound is useful, we set $N=\ln D$ with $N$ the particle number, assume $\|\rho(0)\| = \C O(10^{-N})$ and rewrite
\begin{equation}
 \begin{split}
  \mu_{\rho(0)}&\left\{|S_E^\psi(t)-S_E(t)| > N[\epsilon + g(\epsilon)/N] \right\} \\
  &\le 12 \exp\left(-\frac{4C\epsilon^2}{M^2}10^N\right).
 \end{split}
\end{equation}
Next, let us choose, for example, $\epsilon = 10^{-10}$, which implies
\begin{equation}
 \begin{split}
  & \mu_{\rho(0)}\left\{|S_E^\psi(t)-S_E(t)| > N[10^{-10} + 2.4\cdot 10^{-9}/N] \right\} \\
  &\le 12 \exp\left(-\frac{4C}{M^2}10^{N-20}\right).
 \end{split}
\end{equation}
Now, we see that, say, for a system with $N=30$ particles we have
\begin{equation}
 \mu_{\rho(0)}\left\{|S_E^\psi(t)-S_E(t)| > \frac{N}{10^{10}} \right\} \le 12 \exp\left(-10^{-5}/M^2\right),
\end{equation}
where we used $C = (2304\pi^2)^{-1} > 2.5\cdot10^4$. Finally, let us assume we can distinguish $M=100$ different states with the energy measurement. We then obtain
\begin{equation}
 \mu_{\rho(0)}\left\{|S_E^\psi(t)-S_E(t)| > \frac{N}{10^{10}} \right\} \le 0.000545.
\end{equation}
Thus, on average in only one out of approximately 2000 repetitions of the experiments, one would observe a difference from the ensemble entropy density that is larger than $10^{-10}$.

The above considerations allow us to conclude that, seen on the right extensive scale, significant deviations from the ensemble EP~(\ref{eq EP}) are extremely unlikely to occur in quantum many-body systems unless the initial ensemble $\rho(0)$ lives effectively in a small Hilbert space or the number of measurement outcomes $M$ is unrealistically large. But these are the same conclusions that we also obtained with the example in the main text if one interpretes the heat capacity $C$ as playing the role of $M$: indeed, the heat capacity determines the sensitivity of a system with which it is possible to detect changes of the energy, temperature or entropy (given one of the three quantities as an input). Moreover, note that we here used the same tools and steps as in the main text (we just used a different continuity bound), thus supporting our claim that our argumentation is applicable to a large range of scenarios.


\end{document}